\documentclass[10pt,twocolumn]{article}
\usepackage[T1]{fontenc}
\usepackage[utf8]{inputenc}

\usepackage{graphicx}
\usepackage{algorithmic}
\usepackage{subfigure}
\usepackage{color}
\usepackage{booktabs}
\usepackage{balance}

\usepackage{wrapfig}

\newcommand{\req}[1]{{\bf R#1}}
\newcommand{\nvtree}{NV-tree}
\newcommand{\nvtrees}{\nvtree{}s}
\newcommand{\bplus}{B$^{+}$-tree}
\newcommand{\bpluss}{B$^{+}$-trees}
\newcommand{\btree}{B$^{+}$-tree}
\newcommand{\btrees}{B$^{+}$-trees}

\newcommand{\etal}{et al.}
\newcommand{\citeN}[1]{\cite{#1}}

\begin{document}
\title{Dynamicity and Durability in Scalable Visual Instance Search}


\newlength{\authormax}
\settowidth{\authormax}{xXXXXXXXXXXXXXXXXXXXX}
\author{
Herwig Lejsek\thanks{H. Lejsek and F. H. {\'{A}}smundsson are with Videntifier Technologies, Reykjavík, Iceland. Their work was partly done while at Reykjavik University.}\\
\makebox[\authormax]{Videntifier Technologies, Iceland}\\
herwig@videntifier.com
\and
Bj{\"{o}}rn {\TH}{\'{o}}r J{\'{o}}nsson\thanks{Björn Þór Jónsson is with Reykjavik University, Iceland and IT University of Copenhagen, Denmark. His work was done at Reykjavik University.}\\
\makebox[\authormax]{Reykjav{\'{\i}}k University, Iceland}\\
 ITU Copenhagen, Denmark\\ bjorn@itu.dk
\and
Laurent Amsaleg\\
\makebox[\authormax]{IRISA--CNRS, France}\\
laurent.amsaleg@irisa.fr
\and
Fri{\dh}rik Hei{\dh}ar {\'{A}}smundsson\footnotemark[1]\\
\makebox[\authormax]{Videntifier Technologies, Iceland}\\
fridrik@videntifier.com
}

\date{}
\maketitle

\section*{Abstract}

Visual instance search involves retrieving from a collection of images the ones that contain an instance of a visual query. Systems designed for visual instance search face the major challenge of \textit{scalability}: a collection of a few million images used for instance search typically creates a few billion features that must be indexed. Furthermore, as real image collections grow rapidly, systems must also provide \emph{dynamicity}, i.e., be able to handle on-line insertions while concurrently serving retrieval operations.  \emph{Durability}, which is the ability to recover correctly from software and hardware crashes, is the natural complement of dynamicity. Durability, however, has rarely been integrated within scalable and dynamic high-dimensional indexing solutions.  
This article addresses the issue of dynamicity and durability for scalable indexing  of very large and rapidly growing collections of local features for instance retrieval. 
By extending the \nvtree{}, a 
scalable disk-based high-dimensional index, we show how to implement the ACID properties of transactions which ensure both dynamicity and durability.  We present a detailed performance evaluation of the transactional \nvtree{}: (i) We show that the insertion throughput is excellent despite the overhead for enforcing the ACID properties; (ii) We also show that this transactional index is truly scalable using a standard image benchmark embedded in collections of up to 28.5 billion high-dimensional vectors; the largest single-server evaluations reported in the literature.

\section{Introduction}
\label{sec:introduction}

Visual instance search is the task of retrieving from a database of images the ones that contain an instance of a visual query. It is typically much more challenging than finding images from the database that contain objects belonging  to the same category as the object in the query. If the visual query is an image of a shoe, visual instance search does not try to find images of shoes, which might differ from the query in shape, color or size, but tries to find images of the exact same shoe as the one in the query image. Visual instance search challenges image representations as the features extracted from the images must enable such fine-grained recognition despite variations in viewpoints, scale, position, illumination, etc. Whereas holistic image representations, where each image is mapped to a single high-dimensional vector, are sufficient for coarse-grained similarity retrieval, local features are needed for instance retrieval. 

\subsection{Need for Scalability}

With local features, an image is typically mapped to a large set of high-dimensional vectors, allowing very fine-grained recognition using the multitude of small visual matches between the query instance and the candidate images from the database. Extracting powerful local features from images has been widely studied and many strategies exist to determine (i)~where local features should be extracted in images (\cite{Mikolajczyk2005,DBLP:conf/eccv/NowakJT06}), and (ii)~what information each local feature should 
encode (\cite{DBLP:journals/pami/MikolajczykS05,6248018,Bay2008346}).

All these strategies, however, result in the creation of a very large set of local features per image. In turn, highly efficient high-dimensional indexing techniques are required to quickly return to the user the matching instances. This is an extremely difficult problem as it pushes indexing solutions to their limits in terms of scalability. A system handling a few million images used for instance searches typically has to manage a few billion local features. 
However, most state-of-the-art high-dimensional indexing solutions assume that the feature vector collection can always fit in memory. 

Experience from the data management community and from industry shows that this assumption is not valid, as data will eventually outgrow main memory capacity. Furthermore, with SSDs emerging as an viable middle ground between memory and hard disk drives, the ability to handle data that extends beyond main memory should be reconsidered. 

Industry is  very  concerned with instance search, as there are various real world applications that need such fine-grained image recognition capabilities. Forensics is a domain of choice, where identifying tiny similar visual elements in images is key to mapping out child abuse networks, or establishing links between various terrorism-related visual materials. Very fine-grained instance search is also involved in some copyright enforcement applications, sometimes in conjunction with watermarks. 

\emph{Scalability} of the high-dimensional indexing techniques used in the context of instance search is therefore essential; this is obvious for industry where extremely large image (or video) collections exist, but also for academics as standard visual sets for benchmarking are now quite large: the largest ImageNet set has a few million images, while the YMFCC collection contains 100M images; running visual instance search on such sets would be a good achievement.

\subsection{Need for Dynamicity and for Durability}
But scalability is not the only property that systems for visual instance retrieval must have to be suitable for industry-grade applications. 
First, \emph{dynamicity} is also important. 
Image collections grow (very rapidly) as time goes by, and it is important to ensure that the content-based search engines probe up-to-date collections. Flickr, a popular image sharing site, currently stores about 6 billion high-resolution images, and grows by about 1.5 million every day, while Facebook's  image collection grows by 200 million each 
day.

Very few systems address these dynamicity requirements and most of them can only expand the indexed collection by a complete reconstruction of the index. In the real world, halting a system for complete re-indexing is not an option; instead, new data items must be dynamically inserted to the index while the system is up and running. Even in the recently proposed Lambda Architecture~\cite{lambda:book}, which proposes to separate handling of very recent data from older data, dynamic consistency of index maintenance is desired for the recent data.

Second, resisting failures and enforcing \emph{durability} of the indexed data is very important. Losing the features upon failure or experiencing extended downtime for reconstruction of indices are not acceptable options. Storing the high-dimensional index on disk is thus not only necessary for scalability, but also for the dynamic integrity of the index.

Durability \emph{and} scalability \emph{and} dynamicity have rarely been integrated within high-dimensional indexing solutions. Some contributions address two of these three needs, but, as far as we know, none of them address the three of them simultaneously. 

This article, thus, addresses the issues of dynamicity and durability and scalability jointly implemented within an indexing scheme for the very large and rapidly growing collections of local features that are typically involved in the demanding visual instance search process.

\subsection{Key Requirements}

\label{sec:requ-index-local}

We have identified the following four key requirements 
a high-dimensional indexing solution dedicated to enabling scalable identification of similar visual instances must meet:

\begin{itemize}
\item[\req{1}]	
The 
index must make \textit{efficient use of all available storage resources}, 
main memory, solid-state devices or hard disks. 
\item[\req{2}]	
The 
index must offer \textit{stable query processing performance}, so that it can be used as a component of an industry-scale processing chain. 
\item[\req{3}]	
The 
index must support \emph{dynamic insertion methods} so that the indexed collection can grow while concurrent retrievals are performed.
\item[\req{4}]	
The 
index must support the \textit{ACID properties of transactions} (Atomicity; Consistency; Isolation; and Durability), which guarantee the integrity of index maintenance, as well as correct recovery in the case of system failures.
\end{itemize}

Most state-of-the-art methods, such as product quantization, focus on compressing data into main memory. Whereas these methods often provide some guarantees on quality, they neither consider updates nor provide any guarantees on query processing performance at scale, 
thus failing  with requirements \req{1} to \req{4}. 

The literature describes  the \nvtree{}, an existing  approximate high-dimensional indexing method that is designed from a data management perspective and can thus deal with feature collections that outgrow main memory~\cite{TPAMI,ICMR11}.  When there are more image features that can fit in RAM, the  \nvtree{} guarantees using at most a single disk read per index to get the approximate results. By providing a disk-based query performance guarantee for the  \nvtree{}, requirements \req{1} and \req{2} above are satisfied. 

The \nvtree{}, however, is not a transactional index, as none of the publications describing the \nvtree{} discuss its ability to resist crashes. 
In~\cite{TPAMI}, an insertion procedure is described and evaluated for an early version of the \nvtree{}. Subsequent works, which propose significant improvements to the \nvtree{}, also discuss dynamic maintenance of the index~\cite{ICMR11,ArnarMS}. These publications, however, do not describe insertions in sufficient detail to fully understand how the implementation would support transactions. 
In order to make the \nvtree{} a fully transactional dynamic index, we therefore propose transactional index maintenance procedures, and  show that the resulting extended \nvtree{} supports both \req{3} and \req{4}.

\subsection{Contributions}

This article makes  the following major contributions to the domain of scalable high-dimensional indexing for visual instance search:  
\begin{enumerate}
	\item 
In Section~\ref{sec:acid} we show how to adapt the NV-tree to transactional processing of both insertions and deletions, guaranteeing the well-known ACID properties of transactions. 

	\item
In Section~\ref{sec:dyn:exp} we evaluate insertion performance in this transactional setting and show that insertion throughput is excellent: when the index fits in memory, the index can take full advantage, but even in the disk-bound case each insertion requires only a small fraction of a disk write on average.

	\item 
In Section~\ref{sec:exp} we show that this transactional extension of the \nvtree{} fully works at very large scale. We provide quality measurements of retrievals for the  public Copydays benchmark, embedded in a collection of 28.5 billion ``distracting'' SIFT local features, the largest single-server evaluations reported in the literature.

\end{enumerate}

The technology described in this article is already in use at \emph{Videntifier Technologies}, one of the main players in the forensics arena with technology deployed at such clients as Interpol. Their search engine targets fine-grained visual instance search as it is used for investigations that, for example, aim to dismantle child abuse networks. The search engine can index and identify very fine-grained details in still images and videos from a collection of  150 thousand hours of video, typically scanning videos at 40x real-time speed, and about 700 hours of video material are dynamically inserted to the index every day.

\section{Related Work}
\label{sec:bg}
\label{sec:rel:sota}

This section first gives an overview of the features used for visual instance search. Then, it describes the state-of-the-art in approximate high-dimensional indexing. Finally, it analyses existing indexing methods
%
%
from the point of view of scalability, dynamicity and durability, showing that those few methods that have considered the ACID properties of transactions have not been applied to large-scale collections, and vice versa. 

\subsection{Local Features for Instance Retrieval}
Local features are typically high-dimensional vectors computed in small vicinities of specific areas in images. Such vectors are designed to exhibit some invariance to several image changes, such as: changes of scale; rotations; affine and perspective distortions; crops; changes in illumination; occlusion; etc. Getting local features requires strategies to determine (i)~where such features should be extracted from the images, and (ii)~what information each local feature should encode.

Traditionally, local features have been \emph{hand-crafted} through significant engineering efforts, SIFT being probably the best known such approach~\cite{SIFT99}. Interest points, determined using a Difference of Gaussian, are used to define patches where histograms of gradients are computed, forming the SIFT local features. SIFT has been extended and enhanced by varying the strategy for determining what are the interest points~\cite{Mikolajczyk2005,DBLP:conf/eccv/NowakJT06} and by varying the visual cues around interest points eventually resulting in local features~\cite{DBLP:journals/pami/MikolajczykS05,6248018,Bay2008346}. Overall, up to a few thousands local features can be computed from each and every image, which in turn puts a lot of pressure on storage and on the performance of the indexing schemes. Identifying the images that are similar to a query instance results in multiple nearest neighbor operations over an extremely large index. Attempts to reduce the complexity of retrieval yielded a series of contributions where local features were aggregated into one or very few super vectors. As a result, Sparse coding~\cite{5206757}, Fisher vectors~\cite{DBLP:conf/eccv/PerronninSM10}, and VLAD~\cite{jegou:inria-00633013} were successfully applied to image classification and retrieval, but are too coarse-grained for instance retrieval. 

Recently, however, a new family of local features departs from these historical hand-crafted approaches, as several recent papers rely on \emph{deep learning} methods to learn the size,  location and shape of the regions in images that best work for instance retrieval. Other contributions learn the nature of the local image features that are best suited for instance retrieval~\cite{DBLP:journals/tip/IscenTGJ15,DBLP:conf/cvpr/SalvadorNMS16,DBLP:conf/icassp/JiangZSC16,Tao_2016_CVPR,DBLP:journals/ijmir/AwadKOS17,DBLP:conf/aaai/YuWBY17}. 

Regardless of the techniques that are used to extract local features from images---hand-crafted or based on deep learning---enabling fine-grained instance retrieval always ends up  (a)~identifying many regions in images, and (b)~computing from each region a high dimensional vector. As a consequence, managing a large collection of images (e.g., one million) with an instance retrieval application goal results in managing an extremely large set of local features (e.g., few hundred million descriptors, or even a few billion), which likely outgrows the RAM capacity of standard computers. Some indexing methods address this RAM limit problem, but most contributions do not, as described below.

\subsection{Approximate Nearest Neighbor Algorithms}
Only {\em approximate} high-dimensional indexing solutions remain efficient at very large scale.  Approximate indexing methods trade quality off for response time, and follow three different major directions.  One line of work is based on indexing data clusters such as the hierarchical $k$-means decomposition of the data collection: Voronoi cells are created to partition and store the high-dimensional vectors, and the cells are organized as a multi-level tree to facilitate traversal and improve response time~\cite{DBLP:journals/tc/FukunagaN75}.  Many variants of this basic idea have been proposed (\cite{Clindex02,1641018,4270199,videogoogle,Chierichetti:2007:FNN:1265530.1265545,lostinquantization}).  One algorithm from this category has been extended to cope with collections of up to 43 billion feature vectors, using distributed processing with ``big data'' techniques such as Spark~\cite{gylfi12,gylfi:icmr,gylfi:mmsys}.

A more sophisticated indexing method, still using data clusters at its core, is called product quantization~\cite{jegou:inria-00514462}. 
Product quantization decomposes the high-dimensional space into low-dimensional subspaces that are indexed independently. This produces compact code words representing the vectors that,   together with an asymmetric approximate distance function, exhibit  good performance for a moderate memory footprint. 
Several variants of product quantization have been published (\cite{DBLP:journals/tmm/XioufisPKTV14,DBLP:journals/pami/GeHK014,DBLP:conf/cvpr/KalantidisA14,DBLP:conf/cvpr/HeoLY14}); in particular, Sun \etal~\cite{DBLP:conf/mm/SunWXZ13} proposed an indexing scheme based on product quantization that uses ten computers to fit in memory the 1.5 billion images collection they index. The inverted multi-index by Babenko and Lempitsky~\cite{DBLP:conf/cvpr/BabenkoL12,DBLP:journals/pami/BabenkoL15}, uses product quantization at its core but achieves a much denser subdivision of the space by using multiple inverted indices. The experiments reported in~\cite{DBLP:conf/cvpr/BabenkoL12}, using  the BIGANN dataset~\cite{jegou:inria-00566883} that contains one billion SIFT descriptors, show that the approach can determine short candidate lists with superior recall.  They have recently extended their method for deep learning features~\cite{babenko:deepfeatures}.


A second line of work developed around the idea of hashing. The earliest notable hashing-based method proposed was Locality Sensitive Hashing (LSH)~\cite{LSH99,LSH04}.  Essentially, LSH uses a large number of hashing functions to project high-dimensional vectors onto segmented random lines. At query time, the hash tables are probed with the query vector, and candidates from all these hash tables are then aggregated to find the true neighbors. The performance of such hashing schemes is highly dependent on the quality of the hashing functions. Hence, many approaches have been proposed to improve hashing~\cite{DBLP:conf/nips/WeissTF08,DBLP:conf/cvpr/JainKG08,Tao:2009:QEH:1559845.1559905,DBLP:conf/cvpr/WangKC10,DBLP:journals/prl/PauleveJA10,5767837,DBLP:conf/iccv/JinHLZLCL13}, as well as to reduce the number of hash tables, which in turn diminishes the high storage costs of these tables~\cite{Lv:2007:MLE:1325851.1325958,Joly:2008:PML:1459359.1459388}. 
Tuning hash functions is reported to be a complicated task and some schemes try to automatically adapt to the data distribution~\cite{DBLP:conf/www/BawaCG05}. 

A third approach is based on the idea of a search tree structure. 
The \nvtree{} is one proponent of this group. 
Fagin \etal~\cite{Fagin03} introduced the concept of median rank aggregation. They project the entire data collection on multiple random lines and 
index 
the ranked identifiers of the data points along each line, discarding the actual feature vectors. This ranking turns the high-dimensional vectors into simple sets of values which are inserted to \btrees{}. These \btrees{} are probed at search time, and the nearest neigbors of the query are returned according to their aggregated rankings.  The major drawback of that algorithm is the excessive search across the individual \btrees{}~\cite{Fagin03}.

Tao \etal~\cite{Tao:2009:QEH:1559845.1559905} proposed another  method for accessing high-dimensional data based on \btrees{}, called the locality sentitive B-tree or LSB-tree. The LSB-tree approach inherits some of the properties of LSH, but in addition projects the hashed points onto a Z-order curve. 
Quality guarantees can be enforced using multiple LSB-trees in combination, forming an LSB-forest \cite{Tao:2009:QEH:1559845.1559905,DBLP:journals/tods/TaoYSK10}. 

Muja and Lowe~\cite{flann-pami} proposed, via the FLANN library, a series of high-dimensional indexing techniques based on randomized KD-trees, $k$-means indexing and random projections. Another approach in the category of search trees is the Metric tree~\cite{MetricTree}; a variant named Spill-tree is a tree-structure based on splitting dimensions in a round-robin manner, and introducing (sometimes very significant) overlap in the split dimension to improve retrieval quality~\cite{spillTree}.

\subsection{Scalability, Dynamicity and Durability}
\label{sec:shedd-db-persp}

We now discuss the ability of the various published approaches described above to remain efficient enough at such large scale that disk processing is necessary, as well as their ability to cope with dynamic updates and recover from hardware failures.

\subsubsection{Memory-Oriented Methods}

Overall, most of the high-dimensional indexing techniques for nearest neighbor search disregard disks and focus on main-memory processing. As an example, 
Zäschke \etal~\cite{Zaschke:2014:PSS:2588555.2588564} combine binary patricia-tries
with a multi-dimensional approach similar to quadtrees. The resulting PH-tree is studied with small datasets (at most 100M points) of very low dimensionality (mostly 2D and 3D, but with experiments up to 15D), however, and the memory consumption is comparatively high because the PH-tree embeds the feature vectors in the tree structure.

More advanced methods usually tackle the scalability problem by relying on clever and effective compression mechanisms for the feature vectors---the best example may be the schemes based on product quantization (\cite{jegou:inria-00514462,DBLP:journals/pami/BabenkoL15}). However, there is a limit to the number of images that can be indexed within main memory. According to~\cite{jegou:inria-00514462}, an extremely well optimized indexing scheme based on product quantization needs 32 to 128 bytes per image for near duplicate detection, and several kilobytes for object recognition. 
The inverted multi-index by \citeN{DBLP:journals/pami/BabenkoL15} improves on product quantization, as their most compact proposal uses only 12 bytes per descriptor (4 for the identifier and 8 bytes for information used for improving quality), yet returns better results. Even with the most compact representation, however, it is difficult to index tens of billions of features as a computer with such a large main memory is 
expensive.

Distributed settings can be used to scale to larger collections, but this adds significant complexity and increases the likelihood of failures.  Tao \etal~\cite{spillClustering} were the first to explore this, using the Spill-tree for a collection of 1.5 billion high-dimensional feature vectors, but using the aggregate memory of two thousand workstations were used, presumably having at least a terabyte or two of total main memory.
Sun \etal~\cite{DBLP:conf/mm/SunWXZ13} needed only ten common servers to support real-time search on 1.5 billion images. Each server indexed between 100 and 200 million images with about 60~GB of memory. While scale is addressed to an extent by this work, it is completely main-memory oriented which explains why the aggregated memory of ten servers is needed to fit the collection. Disks are not used, so the index is neither persistent nor durable, and no information on dynamic inserts is given. In fact, if one server fails, the entire system is down and re-indexing the images might be needed. 

Disregarding disks does make it possible for main memory oriented high-dimensional indexing schemes to improve the quality of their results by analyzing a large amount of data in memory at search time; this would be much too costly if disks were involved due to the significant number of (mostly random) disk reads that would be required. 
%
But disregarding disks precludes these contributions from coping with failures and recovering from crashes. In addition, the ability of most indexing techniques to cope with dynamic inserts remains a question, in particular at large scale. 


\subsubsection{Disk-Oriented Methods}

Many variants of the original R-trees and KD-trees do take disks into account and support dynamic inserts. Concurrency control algorithms have been developed for these two indexing schemes and they can be made fault tolerant by implementing the write-ahead-logging protocol. These two approaches, however, are known to perform poorly when indexing large collections of high-dimensional data. 

Multiple randomized KD-trees~\cite{flann-pami} cope better with scale. The datasets they used were significant, both when a single server was used and when distributed search across multiple machines was used to cope with the 80 million tiny images of~\cite{DBLP:journals/pami/TorralbaFF08}. However, it has not yet been demonstrated that randomized KD-trees can handle collections containing a billion vectors or more. The API in the FLANN library for randomized KD-trees only allows for bulk-loading the index, with no suggestion that dynamic inserts are supported. The index can be pushed to disk and later read back, but no comment on recovery is provided~\cite{flann-pami}. 

The LSB-tree approach by \cite{Tao:2009:QEH:1559845.1559905} clearly considers disks as it is implemented within a relational database engine. This work naturally copes with dynamic updates and resists failures. This approach, however, has only been tested using very small datasets---the largest one including only 100,000 points---and it is hard to predict how well it might behave at much higher scales. Among the expected obstacles to scalability is the use of the Z-order curve which is known to have a poor order-preserving behavior (other, more complicated, space-filling curves could perform somewhat better). Overall, many LSB-trees must be used to enforce quality guarantees, which in turn requires performing many disk operations, which eventually endangers scalability. The experiments reported by~\cite{Tao:2009:QEH:1559845.1559905} show that about 10 disk reads are needed for a single LSB-tree, even with their small collections, while a few hundred disk reads are required for the corresponding LSB-forest.

\begin{figure*}[t!]
	\centering
	\includegraphics[angle=-90,width=1.5\columnwidth]{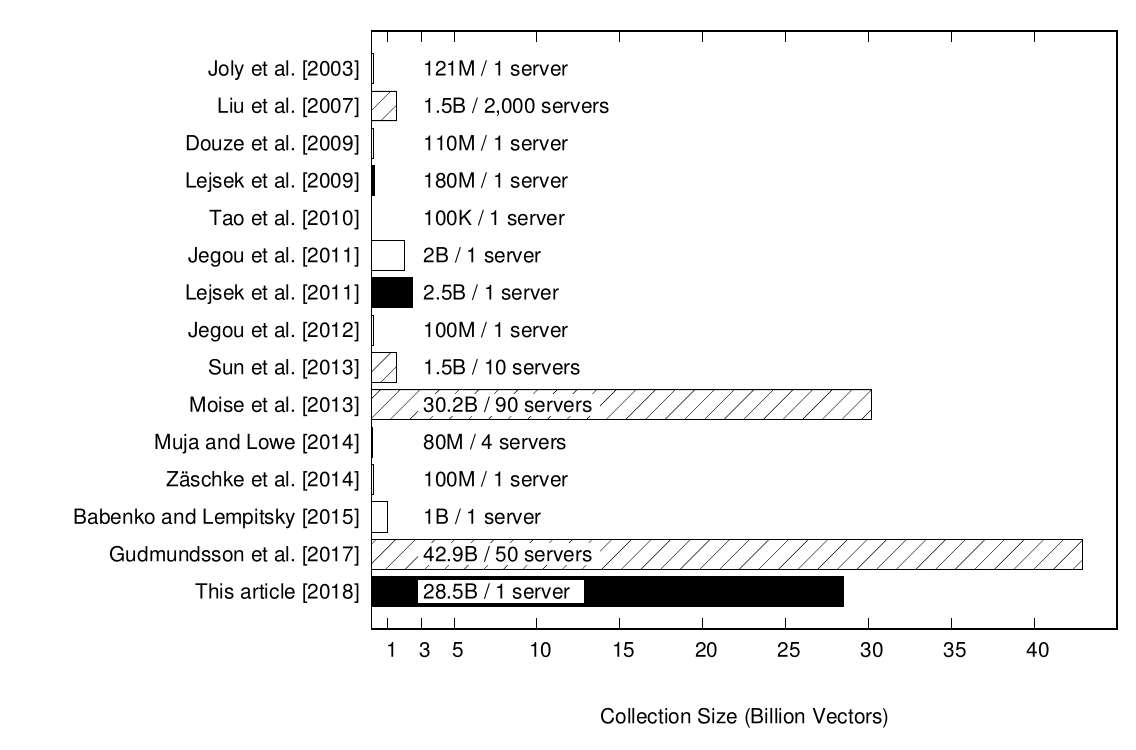}
	\caption{Comparing scales of experiments in the literature. Black bars represent results using the \nvtree{}. Shaded bars represent multi-server configurations.}
	\label{fig:scalecomparison}
\end{figure*}

The only method which has addressed scale similar to that reported in this article, is the DeCP algorithm~\cite{gylfi:icmr,gylfi:mmsys}. That work focuses on a very simple disk-based clustering method 
but uses a large number of workstations to index the collection and answer queries. While the scale of the experiments is indeed impressive, the philosophy of the system is quite different: it is dedicated to processing large batches of queries and cannot run interactively. Furthermore, that work has not addressed the dynamicity and durability requirements.

 \subsection{Summary}


We conclude this review by summarising the scale of the experiments found in the literature described above.  In addition to the work proposed here,  we include 14 contributions where either the proposed high-dimensional indexing schemes or the scale of the associated experiments contributed significantly to advancing the state of the art.  
Figure~\ref{fig:scalecomparison} shows these contributions along a timeline; as the figure shows our experiments use by far the largest feature collection ever reported in a single-server setting. 

\section{The \nvtree}
\label{sec:nvtree}
\label{sec:nv}

The \nvtree{}~\cite{TPAMI, ICMR11} is a disk-based high-dimensional index. It builds upon a combination of projections of data points to lines and partitioning of the projected space. By repeating the process of projecting and partitioning, data is separated into small partitions which can be easily fetched from disk with a single read, and which are likely to contain all the close neighbors in the collection.  
We briefly describe  the NV-tree creation process, its search procedure, its dynamic insert process and and then enumerate some of its salient properties.

\subsection{Index Creation}
\label{sec:tree:algorithm}
Overall, an \nvtree{} is a tree index consisting of: a)~a hierarchy of small {\em inner nodes}, which guide the vector search to the appropriate leaf node; and b)~larger {\em leaf nodes}, which contain references to actual vectors. The leaf nodes are  further organised into {\em leaf-groups} that are disk I/O units, as described below.

When  tree construction starts, all vectors from the collection are first projected onto a single projection line through the high-dimensional space (\cite{TPAMI} discusses projection line selection strategies).  
The projected values are then partitioned in 4 to 8 partitions based on their position on the projection line. 
Information about the partitions, such as the partition borders along the projection line, forms the first inner node of the tree---the root of the tree.
To build the subsequent levels of the \nvtree{}, this process of projecting and partitioning is repeated recursively for each and every partition, using a new projection line for each partition, thus creating the hierarchy of smaller and smaller partitions represented by the inner nodes.

At the upper levels of the tree, with large partitions, the partitioning strategy assigns equal distance between partition boundaries at each level of the tree. 
The partitioning strategy changes when the vectors in the partition fit within 6$\times$6 leaf nodes of 4~KB each. In this case, all the vectors from that partition are partitioned into a {\em leaf-group} made of (up to) 6 inner nodes, each containing (up to) 6 leaves. In this leaf-group, partitioning is done according to an equal cardinality criterion (instead of an equal distance criterion).  Finally, for each leaf node, projection along a final random line gives the order of the vector identifiers and the ordered identifiers are written to disk. It is important to note that the vectors themselves are \emph{not} stored; only their identifiers.

Indexing a collection of high-dimensional vectors with an \nvtree{} thus creates a tree of nodes keeping track of information about projection lines and partition boundaries. All the branches of the tree end with leaf-groups with (up to) 36 leaf nodes, which in turn store the vector identifiers.

\subsection{Nearest Neighbor Retrieval}
\label{sec:tree:search}

During query processing, the search first traverses the hierarchy of inner nodes of the \nvtree{}. At each level of the tree, the query vector is projected to the projection line associated with the current node. The search is then directed to the sub-partition with center-point closest to the projection of the query vector until the search reaches a leaf-group, which is then fully fetched into RAM, possibly causing one single disk I/O. Within that leaf-group, the two nodes with center-point closest to the projection of the query vector are  identified. The best two leaves from each of these two nodes are then scanned in order to form the final set of approximate nearest neighbors, with their rank depending on their proximity to the last projection of the query vector.  The details of this process can be found in~\cite{ICMR11}.

While the  \nvtree{} is stored on disk, the hierarchy of inner nodes is read into memory once query processing starts, and remains fixed in memory. The larger leaf nodes, on the other hand, are read dynamically into memory as they are referenced. If the \nvtree{} fits into memory, the leaf nodes  remain in memory and disk processing is avoided, but otherwise the buffer manager of the operating system may remove some leaf nodes from memory. 

\subsection{Insertions}
\label{sec:insertions}
\label{sec:dyn:ins}

Insertion to \nvtree{} leaf nodes proceeds as follows. First, the leaf node where to insert a new vector identifier is identified. The position within that leaf is also determined and the insert is performed if the leaf is not full. As for most dynamic data structures, leaf nodes at index creation time are not filled completely (they are between 50\% and 85\% full, and about 70\% full on average) in order to leave space for such insertions. A filled  leaf node must be split in order to provide more storage capacity within the tree. During a split operation, a leaf-group is considered as a unit, and all the features of the leaf-group are re-organized using the same process as during index construction. In particular, when the size of the leaf-group exceeds the capacity of $6 \times 6$ leaf nodes, the group is split into 4 to 8 new leaf-groups, depending on the distribution of the features.

During the leaf-group re-organization, new projection lines may be chosen for the internal nodes, and each new leaf will have a new projection line.  As leaf nodes only contain vector identifiers, the vectors must be retrieved from disk for re-projection. In~\cite{ICMR16}, it is shown that the most efficient option for handling re-projections is to maintain an independent feature database for each NV-tree, organized in the same manner as the leaf-groups, which allows directly reading the relevant features.

\subsection{Properties of \nvtree{}s}

The experiments and analysis of \citeN{ICMR11} show that the \nvtree{} indexing scheme has the following properties:
\begin{itemize}
	\item \emph{Random Projections and Ranking:} 
          The \nvtree{} uses random projections turning multi-dimensional vectors into single-dimensional values indexed by \bpluss{}. Efficient implementations of dynamic \bpluss{} are well known. The \nvtree{} does not fetch full vectors from disks to subsequently compute distances. In contrast, ranking is used, which basically amounts to scanning a list.
	
	\item \emph{Single Read Performance Guarantee:} 
	In the \nvtree{}, leaf-groups have a fixed size. Therefore, the \nvtree{} guarantees query processing time of a single read regardless of the size of the vector collection.  Larger collections need deeper \nvtree{}s but the intermediate nodes fit easily in memory and tree traversal cost is negligible.

	\item \emph{Compact Data Structure:} 
	The \nvtree{} stores in its index the identifiers of the vectors, not the vectors themselves. This amounts to about 6 bytes of storage per vector on average.  The \nvtree{}  is thus a very compact data structure.  Compactness is desirable as it maximizes the chances of fitting the tree in memory, thus avoiding disk operations.

	\item \emph{Consolidated Result:} 
	Random projections  produce numerous false positives that can be almost all eliminated by an ensemble approach. Aggregating the results from a few \nvtree{}s, which are built independently over the same collection, dramatically improve result quality.

\end{itemize}

\subsection{Discussion}
The \nvtree{} is however not a transactional index. None of the publications describing this index provide any solid discussion on its ability to resist crashes. By construction, it is obvious that some of the indexed data is preserved on disks, allowing to possibly recover from a main memory failure by re-loading parts of the index and resuming work. \emph{Durability}, however, is by no means guaranteed as the conditions for propagating the updates that are in RAM to disks are unspecified. Furthermore, in case of a devastating media crash, no mechanism allowing to reconstruct the lost storage is discussed. Last, the specifications for handling dynamic inserts are not clear enough, leaving it unclear whether or not multiple inserts can be performed simultaneously to multiple reads. It is thus impossible to positively assert that the \nvtree{} satisfies \req{3} \emph{and} \req{4}.

The next section fills these gaps and specifies a model for the concurrent execution of updates and retrievals. It also specifies the conditions for ensuring the durability of the indexed data, as well as the updates. Finally, it discusses mechanisms to facilitate recovery in the event of disk crashes.

\section{Transactional \nvtree}
\label{sec:dynamic}
\label{sec:dyn}
\label{sec:acid}

Large collections of media objects, and the corresponding collections of high-dimensional vectors, are typically dynamic and require efficient insertions.  For the typical web-scale application of visual instance search, however, it is safe to assume that (a)~updates are made centrally, and (b)~that throughput is more important for this update thread than response time.  For these applications, it is feasible to batch insertions such that only one insertion thread is running at each time, which  simplifies the implementation of the insertion process. Of course, however, insertions and searches must run concurrently. 

This section focuses on insertions to the \nvtree{} index. We outline the insertion operation, then describe enforcement of the ACID properties of transactions (Atomicity; Consistency; Isolation; and Durability), and finally consider the correctness and performance of the proposed methods. Note that while deletions will be rare in practice, they can be implemented using techniques very similar to those implementing insertions. We therefore  briefly describe the differences for deletions, where appropriate. Updates are implemented as feature deletions followed by feature insertions. 

\subsection{Enforcement of ACID Properties}
\label{sec:large}
\label{sec:dyn:acid}

Due to serialization of inserts, two insertion transactions will never conflict, which means that a simple locking mechanism based on tree-traversals is sufficient to enforce {\em isolation}. 
Because insertions are never aborted and they never deadlock, ensuring {\em atomicity} is only needed when the system crashes.  Furthermore, since at most one insert transaction is running concurrently, enforcing {\em durability} is greatly simplified.  Finally, since there are no constraints on the vectors, as such, the notion of {\em consistency} simply implies that the results always reflect the status after the last committed transaction.

We start by considering isolation and consistency for a single \nvtree{}. We then consider atomicity and durability, before addressing some practical issues relating to using  multiple \nvtree{}s.  

\subsubsection{Isolation and Consistency}
\label{sec:dyn:acid:iso}

Isolation is implemented by adapting a standard locking algorithm from the \bplus{} literature~\cite{grayreuter,btrees}.  A search thread starts by obtaining a read lock on the root of the \nvtree{}. Before accessing a child node, the thread must obtain a read lock on that node. At that point, the lock on the parent can be released. Finally, the leaf-group selected for retrieval is locked and only released after all necessary identifiers have been retrieved from the leaves.  Note that locks are implemented using {pthread} mutexes; each internal node contains the mutexes for all its children and the leaf-groups are locked as a unit since they are treated as a unit during both retrieval and node splits. As the overhead of obtaining mutexes is low, locking is always activated.

The insertion process uses the same locking mechanism, except that finally an exclusive lock is acquired for the leaf-group, preventing concurrent insertions into that leaf-group, as well as concurrent retrieval from the leaf-group.  In the case of a leaf-group split, a new internal node is created pointing to all the newly created leaf-groups; the lock on the original leaf-group is sufficient to protect the modification of the parent node.

Since each query or insertion transaction needs to access multiple trees multiple times, it is necessary, however, to consider the overall interaction between search and insertion transactions.  Recall that insertion transactions are serialized; they are therefore assigned with ever-increasing transaction identifiers (TIDs) that are logged with each inserted vector.  Isolation is then enforced by omitting from the query result vectors with transaction identifiers larger than that of the last transaction that committed before the search started; this also guarantees consistency of the result.

Deletions are implemented in the same manner as insertions, except that a list of deleted media items is maintained to avoid returning partially deleted items; when all  feature vectors from a media item have been deleted, it can be removed from this list. 

\subsubsection{Atomicity and Durability}
\label{sec:large:thread:atomicity}
\label{sec:dyn:acid:ato}
\label{sec:dyn:acid:dura}

For atomicity and durability, we adopt the standard write-ahead logging (WAL) protocol~\cite{GrayWALogging,Mohan92aries}. The WAL protocol uses a transaction log (or write-ahead log) which contains sufficient information to recover in case of failures.  
The WAL protocol has two rules  to ensure the correctness of transactions:
\begin{enumerate}
\item The log entry for any modification must be written to disk before the modified data is written to disk.
\item All log entries for a transaction must be written to disk before the transaction can be committed.
\end{enumerate}
The first rule---sometimes called the {\em undo} rule---ensures that any change that is written to the disk before it is committed can later be removed from the database, thus supporting atomicity.  The second rule---the {\em redo} rule---ensures that any committed changes can be redone in case of crashes, thus supporting durability.

Each split can result in a large number of disk operations and splits are therefore heavily buffered. 
It is best for performance to manage multiple log files where log records can be appended independently and in parallel. There is one log file per \nvtree{} plus a global log file for the correctness of the overall recovery process.

The recovery manager uses regular {\em checkpoints} to facilitate efficient recovery.  
During recovery, the latest checkpoint file is first read and the status at the time of the checkpoint is adopted for the internal nodes, the leaf nodes and the leaf-group DB. Then the split operations are retrieved from the index log file, and those split operations that were performed due to committed transactions are re-played on the internal structure, while other split operations are ignored. At this point, the internal structure is correct, as of the time of the crash, but vectors may be incorrectly included and/or missing. Next, therefore, vectors that belonged to uncommitted transactions, but made it to the leaf nodes of the \nvtree{} are removed; note that no such vectors are ever found in the leaf-group DB, because they are only added to the leaf-group buffer when the transaction is ready to commit and the checkpoint is only written after commit. Finally, the vector collection log file is used to re-insert the committed vectors that did not make it to disk, both to the \nvtree{} and the leaf-group DB, taking care to avoid re-insertion to the split leaves.

Note that since the insertion operations are serialized and do not conflict, the undo and redo phases can be performed in any order.  Since vector removal requires moving other vector identifiers in the leaves, however, it makes sense to do that before inserting new identifiers that would subsequently need to be moved.

\subsubsection{Practical Issues with Multiple \nvtree{}s}
\label{sec:large:thread:bottlenecks}
\label{sec:dyn:acid:multi}

When inserting to multiple \nvtrees, each tree should preferably be located on a separate hard drive (as should the log files) so that the full write-back capacity of the disks can be used for the leaf-group DB thread.  In order to fully use the capacity of the disks, however, it is important to decouple the insertion process (as well as logging and checkpointing) for each \nvtree.
Each \nvtree{} can thus be inserting from a different transaction, but they must all process the transactions in the same order. Since transactions may progress differently across different trees, more than one uncommitted transaction may have inserted vectors to some trees before a crash. Due to the ordering of transactions, however, the last \nvtree{} to finish a transaction decides the commit time and transactions will therefore commit in the same order, and all the techniques described above are unaffected by this change.  Using decoupling, disk utilization was improved from about 40\% up to 75\% to 80\%, without violating the previously described ACID properties.

\subsection{Correctness and Recovery Performance}
\label{sec:dyn:acid:proof}

Since our techniques are built on standard building blocks from the database literature, which have been shown to enforce the ACID properties, a formal proof of correctness is beyond the scope of this paper.
In the following, however, we give a brief outline of how such a proof would be structured.

A sufficient condition for enforcing isolation is {\em serializability}. 
Recall that we assume that insertion transactions stem from a single, serialized thread. Then the only conflicts that can arise are between this single insertion transaction and the (potentially many) retrieval transactions. As we use standard \btree{} locking for the data structure consistency, which is known to enforce serializability, and further ensure that retrieval transactions can only see insertions from insertion transactions that committed before the retrieval started, isolation is fully enforced.  And, as discussed above, since there are no constraints on the vectors, isolation is sufficient to enforce consistency for the class of applications considered here.

By  definition, the WAL protocol enforces both atomicity and durability.  The fact that the log is stored in multiple files does not change this property, as long as sufficient information is stored in the log entries to redo operations in the correct order.  As described in detail above, the recovery operations have been carefully ordered to ensure correctness. The proposed method therefore enforces both atomicity and durability.

Proving the correctness of the {\em implementation} of our method, on the other hand, is of course extremely difficult, if at all possible.  The implementation has been tested very methodically, however, by pausing operations in certain places and crashing the computer; in all cases has recovery been successful.  The recovery performance depends on the frequency of the checkpoints, but with reasonable checkpoint frequency the database is always fully recovered within a matter of minutes even with very large collections.

\section{Performance of Index Maintenance}
\label{sec:dyn:exp}

In this section we investigate the performance of dynamic inserts, while guaranteeing ACID properties, as described above. As the index experiences splits upon inserts, it is also important to verify that the evolution of the data structure does not impact the ability of the \nvtree{} to correctly identify nearest neighbors. We first discuss insertion throughput and then result quality.

\subsection{Experimental Setup}

This experiment was designed to show the two interesting cases that govern the performance of inserts. First, when the index fits in RAM, inserts are done in memory and later asynchronously pushed to disks, resulting in excellent performance. The second case arises when the index is larger than memory. In this case, loading the affected data pages from disk may be required, which is not only slow but also interferes with writing back updated pages. We therefore expect this second case to show much worse performance. 

To illustrate these two cases, we used a machine with only 32GB of main memory. We used a small subset of a very large collection of images from Flickr (see~\ref{distractors}) to first compute 36 million SIFT vectors~\cite{SIFT04}. We then indexed these 36 million vectors with three \nvtrees. This is a tiny collection which can be indexed very quickly, and the resulting \nvtree{}s together occupy slightly more than 500MB. We then ran sequences of 1,000 insertion transactions. Each transaction is inserting 100,000 new vectors into the three \nvtrees{}, which means that each sequence of insertion inserts 100 million new vectors. We then observed the time it takes for each sequence to complete. We repeated this process and ran multiple sequences until each of the \nvtrees{} contained nearly 2.5 billion vectors, occupying about 328~GB each.  

\begin{figure}
	\centering
	\includegraphics[width=.9\columnwidth]{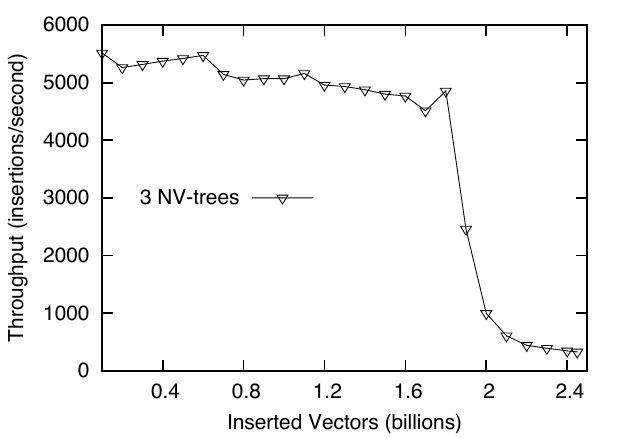}
	\caption{Insertion throughput (dynamic collection; three \nvtree{}s; six hard disks; 32 GB of main memory).}
	\label{fig:inserts}
\end{figure}

\subsection{Insertion Throughput}

Figure~\ref{fig:inserts} shows the evolution of the insertion throughput (measured by vectors inserted per second) for the duration of this workload.  In the beginning of this workload, all three \nvtrees{} fit into main memory and the throughput is excellent, around five thousand vectors per second. After running 18 such transactions, thus inserting 1.8 billion vectors, the 3 \nvtree{}s no longer fit in main memory. After that point, Figure~\ref{fig:inserts} clearly shows the insert behavior corresponding to the second case discussed above, where the rate of inserts slows down significantly due to conflicting disk operations.  It should be noted, however, that with throughput of 500 vectors per second, each insertion only takes 2 ms., which is significantly less than one disk operation per insertion, even though the descriptors are inserted to three \nvtrees{} simultaneously.

The most important aspect of this experiment is not the reduced performance of inserts after 1.8 billion vectors have been inserted and the index no longer fits in memory; by adding more memory, larger indices can be stored in RAM.
Rather it is the fact that even when the collection no longer fits in memory, and must be stored on traditional HDDs, dynamic maintenance of the index is still  possible as the insertion throughput degrades gracefully.  Performance of index maintenance could be vastly improved if SSDs were used instead of HDDs. SSDs are nowadays a viable alternative to HDDs:  Their capacity increases very quickly, their cost decreases as well. SSDs outperform by far HDDs, and the conflicting operations (random reads and writes) would not have such a negative impact of the performance of inserts. Note that using SSDs instead of HDDs would also improve the performance of the retrievals as read operations on electronic secondary storage are much faster than magnetic ones -- this is discussed below.

\subsection{Retrieval Quality at a Moderate Scale}

To evaluate the query performance of the \nvtree{}, we borrow the ground truth defined by \citeN{TPAMI}. A sequential scan was used to determine the 1,000 nearest neighbors of 500,000 query vectors, all coming from a very large collection of SIFTs. The resulting 500M neighbors were then analysed to identify 248,212 vectors as being meaningful nearest neighbors of the query points (as defined by \citeN{beyer99when} and \citeN{Hinneburg00}).

To reuse that workload, we included these 248,212 vectors in the database of 36 million other vectors used previously. Once this database was created, we ran the same 500,000 queries as in \citeN{TPAMI} and computed their recall, i.e.,  we counted how many of these 248,212 ground truth vectors were found. We repeated that same workload after every insertion transaction (of 100 million vectors), to observe how the quality of the answers evolves as the database grows.

\begin{figure}
	\centering
	\includegraphics[width=.9\columnwidth]{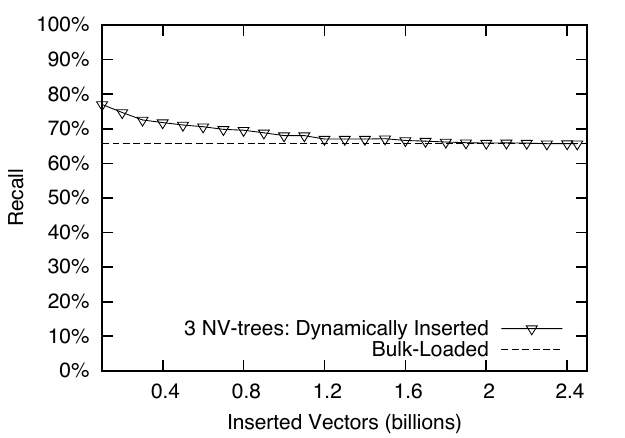}
	\caption{Insertion quality (dynamic collection; three \nvtree{}s).}
	\label{fig:insertQuality}
\end{figure}

Figure~\ref{fig:insertQuality} plots the recall percentage from the 500,000 queries described above, as the collection grows in size. The figure shows a configuration where the results are aggregated from three \nvtree{}s. As the figure shows, recall drops slowly as the collection grows, which was expected. For comparison, the figure also contains a dashed line indicating the result quality when the \nvtrees{} for the 2.5 billion vectors are constructed from scratch via bulk loading. As the figure shows, the results for the dynamically created \nvtrees{} and the bulk-loaded \nvtrees{} are identical, meaning that dynamicity has no impact on result quality.

\subsection{Retrieval Quality at Large Scale}
\label{distractors}
The previous experiments demonstrated that dynamicity has no impact on quality at moderate scales. The same applies at larger scale as we show here. At larger scale, quality degrades slowly as more and more distracting vectors get in the way of the retrieval process. To evidence this phenomenon, we have repeated the previous experiment and drown the workload described above within vector sets of varying cardinalities to distract the search. These sets of {\em distracting vectors} have been created by extracting SIFT features from images randomly downloaded from Flickr between 2009 and 2011. The downloading process rejected  images smaller than 100x100 pixels and also used MD5 signatures to reject exact duplicates of any previously downloaded images. We have gathered almost 30 million such images, and the  resulting distractor sets contain about 30 million vectors, 180 million, 300 million, 2.5 billion, 3 billion and 28.5 billion vectors, respectively.

For that experiment, we used a Dell r710 machine that has two Intel X5650 2.67~Ghz CPUs. Each CPU has 12~MB of L3 cache that is shared by 6 actual and 6 virtual cores. There are therefore 24 cores; in particular 12 are actual processing units.  The RAM consists of 18x8~GB 800~Mhz RDIMMs chips for a total of 144~GB. That machine is connected to a NAS 3070 storage system from NetApp, offering about 100TB of magnetic disk space in a RAID configuration. We ran the experiments using a single core, however, using three \nvtrees{} which are probed one after the other; no parallelism is enforced in our experiments while this could be trivially done. 

\begin{table*}[t!]
  \renewcommand{\arraystretch}{1.3}
  \caption{Distracting vector collections}
  \label{tab:overviewOfDataSets2}
  \centering
  \begin{tabular}{rrrr}
    \hline
      & \multicolumn{2}{c}{Collection Size} & \multicolumn{1}{c}{\nvtree{} size} \\
    \multicolumn{1}{c}{Collection}   & (SIFT vectors) &  (on disk) & (one tree, on disk) \\
  \hline
    30M~~~ & 28,799,690 & 3.8 GB & 180 MB~~~~~~\\
    180M~~~ & 179,443,881 & 23.6 GB  & 1 GB~~~~~~\\
    300M~~~ & 305,443,749 & 40.3 GB & 1.9 GB~~~~~~\\
    2.5B~~~ & 2,485,568,191 & 328 GB & 14 GB~~~~~~\\
    3B~~~ & 3,040,856,472 & 401 GB & 17 GB~~~~~~\\
    28.5B~~~ & 28,484,904,924 & 3.7 TB & 162 GB~~~~~~\\
    \hline
  \end{tabular}
\end{table*}
With the 30M collection, recall is 79\% using three \nvtree{}s. With the 28.5B collection, recall is lower but remains remarkably good given the size of the distracting collection: 58\%.

We now turn to the retrieval performance. We measured the response time of each individual query as well as the throughput of the system, determining the number of query vectors it can process per second. 
As the results are highly dependent on hardware, and memory size in particular, we focus on the key retrieval performance elements: the dominating costs related to the CPU consumption and main memory latency when the \nvtree{} indices fit in main memory; and the performance of disk reads when the indices can no longer fit within memory.

Recall that the main memory of our server was 144~GB, which means that all the leaves of three \nvtree{}s can fit into memory for all collections except the 28.5B collection. When the various indices entirely fit in main memory, then answering each query vector is extremely fast. It takes a fraction of a millisecond to process one vector against one \nvtree{}, and the throughput we observed ranged from 2,000 to 3,000 query vectors per second per tree. 

It should be noted, however, that this throughput can be achieved only once each \nvtree{} index entirely resides in main memory, that is, once all its leaves are in RAM. The leaves can be purposely loaded to memory before running queries, or loaded as a consequence of the querying process. In the latter case, the first queries are slow as they need to fetch data from disks, while subsequent queries are faster as they find more and more likely the data they need in memory, loaded by previous queries.

When using the 28.5B collection, on the other hand, the main memory can not fit even all the leaves of one \nvtree. Each query vector is likely to access a different random part of the index and no buffering policy copes with such demanding access patterns, meaning that the system must retrieve data from disks for almost every query vector. The response time is therefore much larger.

The duration of each I/O varies but typically is within a range of 5 to 20 milliseconds. I/Os are very random and it is extremely complicated to precisely know how they are handled by the NAS NetApp server. It serves many users in parallel, has various level of caches that we can neither control nor observe, and stripes the data across its disks in an opaque manner. Overall, however, about 50 query vectors could be processed per second per tree, as the \nvtree{} meets the design criterion of one disk read per query.

\section{Scalability of the transactional \nvtree{}}

\label{sec:exp-apps}
\label{sec:exp}

We now describe a large scale benchmark experiment in order to demonstrate the scalability of the transactional \nvtree{}. We use the Copydays benchmark embedded in collections of up to 28.5B  high-dimensional descriptors---the largest single-server experiments ever reported in the literature.  We first discuss the experimental protocol and the image collections and queries used, before reporting the results. 

\subsection{Experimental Protocol}

The focus of the \nvtree{} is on supporting visual instance search. Unfortunately, however, no large scale instance search benchmarks exist in the literature, as (i)~generating the ground truth for instance search is difficult, and (ii)~the machine learning techniques used today to determine which regions should be identified and what local info to capture in features are not yet sufficiently scalable to generate collections consisting of tens of billions of features. We must therefore simulate instance search with other existing technologies.

We observe that regardless of the nature of the final feature vectors that are for instance search, this domain will always demand many local features for each image, and querying will boil down to running multiple $k$-NN queries and consolidating the result afterwards into a single reply. The best representative for the visual instance search domain that we are aware of is using SIFT features:  many local descriptors are generated for each database and query image and result consolidation can be done via simple voting schemes.

\subsection{Image Dataset and Ground Truth}
\label{sec:image-dataset-ground}

In order to evaluate the performance of the \nvtree{}, we have therefore adopted a traditional fine-grained quasi-copy paradigm. We use the well known and public Copydays benchmark~\cite{Douze:2009:EGD:1646396.1646421} where predefined image transformations have been applied to a particular collection of images in order to obtain a set of query images. We then ``drown'' the original images, used to create the transformed quasi-copies, within a large collection of random images which play the role of ``distracting'' the search. 
All pictures used in our experiments were resized such that their longer edge is 512 pixels long.

The transformed query images are then evaluated against the indexed collection and the location of the original image in the final result list is noted. When the first image in the ranked result list is the original image, the answer is considered correct; if the first image in the ranked list is \emph{not} the original image, then the system is said to \emph{fail}, even if that image turns out to be second in the ranked list. For each transformation, 100\% success means that all the ground truth images were at the top of the corresponding ranked lists in the result set.

Copydays contains 157 original images. Three families of transformations have been applied, resulting in 3,055 quasi-copies in total.  Some of the 229 manual transformations are particularly difficult to find since they generate quasi-copies that are visually extremely different from their original counterparts. For example, Figure~\ref{fig:exampleCopyDays} shows two original images, \subref{fig:exampleCopyDays-a} and~\subref{fig:exampleCopyDays-c}, and one strong (manually created) variant, in~\subref{fig:exampleCopyDays-b} and~\subref{fig:exampleCopyDays-d} respectively (note that the relative sizes of the originals and their variants is preserved).

\newlength{\copydayslength}
\setlength{\copydayslength}{1.130\columnwidth}

\begin{figure}[t!]%
\centering
\subfigure[][]{%
\label{fig:exampleCopyDays-a}%
\includegraphics[width=0.30\copydayslength]{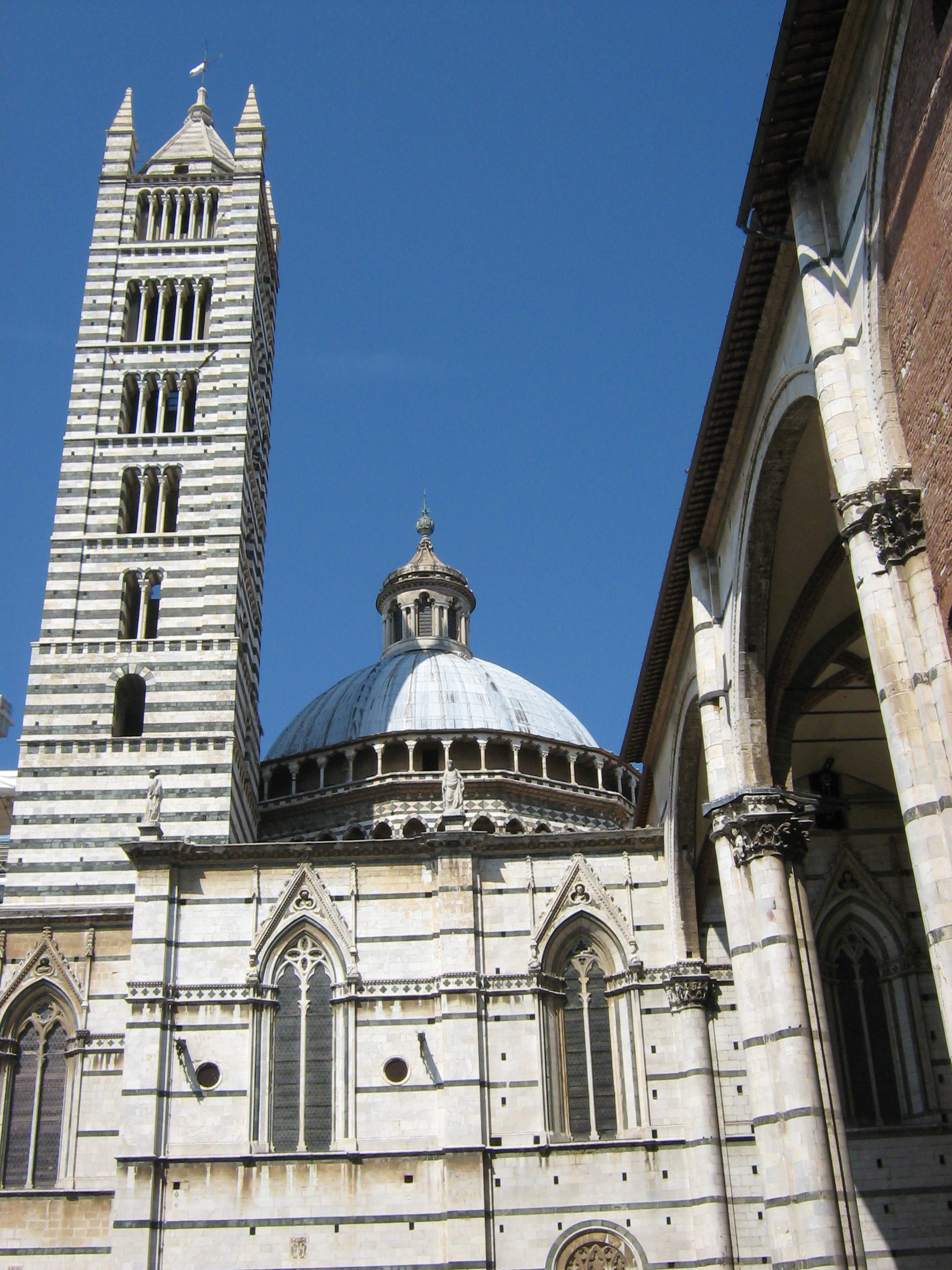}}%
\hspace{8pt}%
\subfigure[][]{%
\label{fig:exampleCopyDays-b}%
\includegraphics[width=0.14\copydayslength]{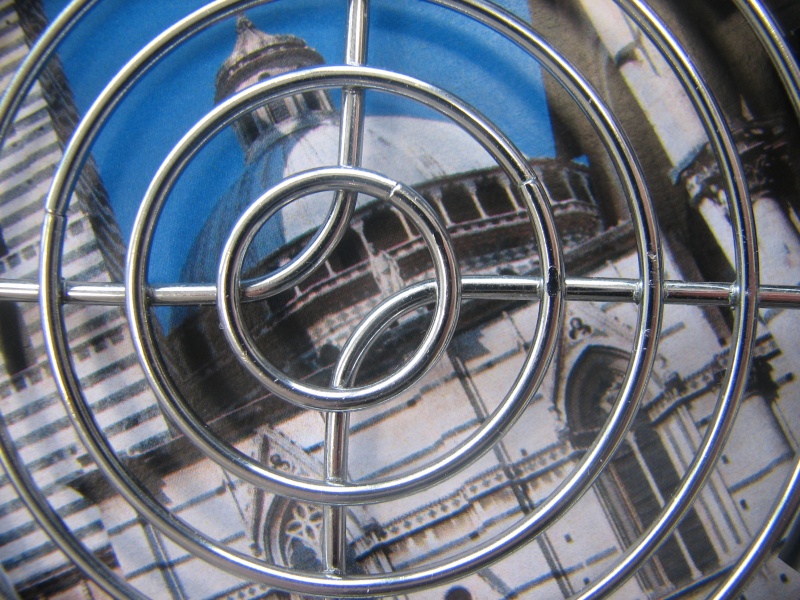}}\\
\subfigure[][]{%
\label{fig:exampleCopyDays-c}%
\includegraphics[width=0.30\copydayslength]{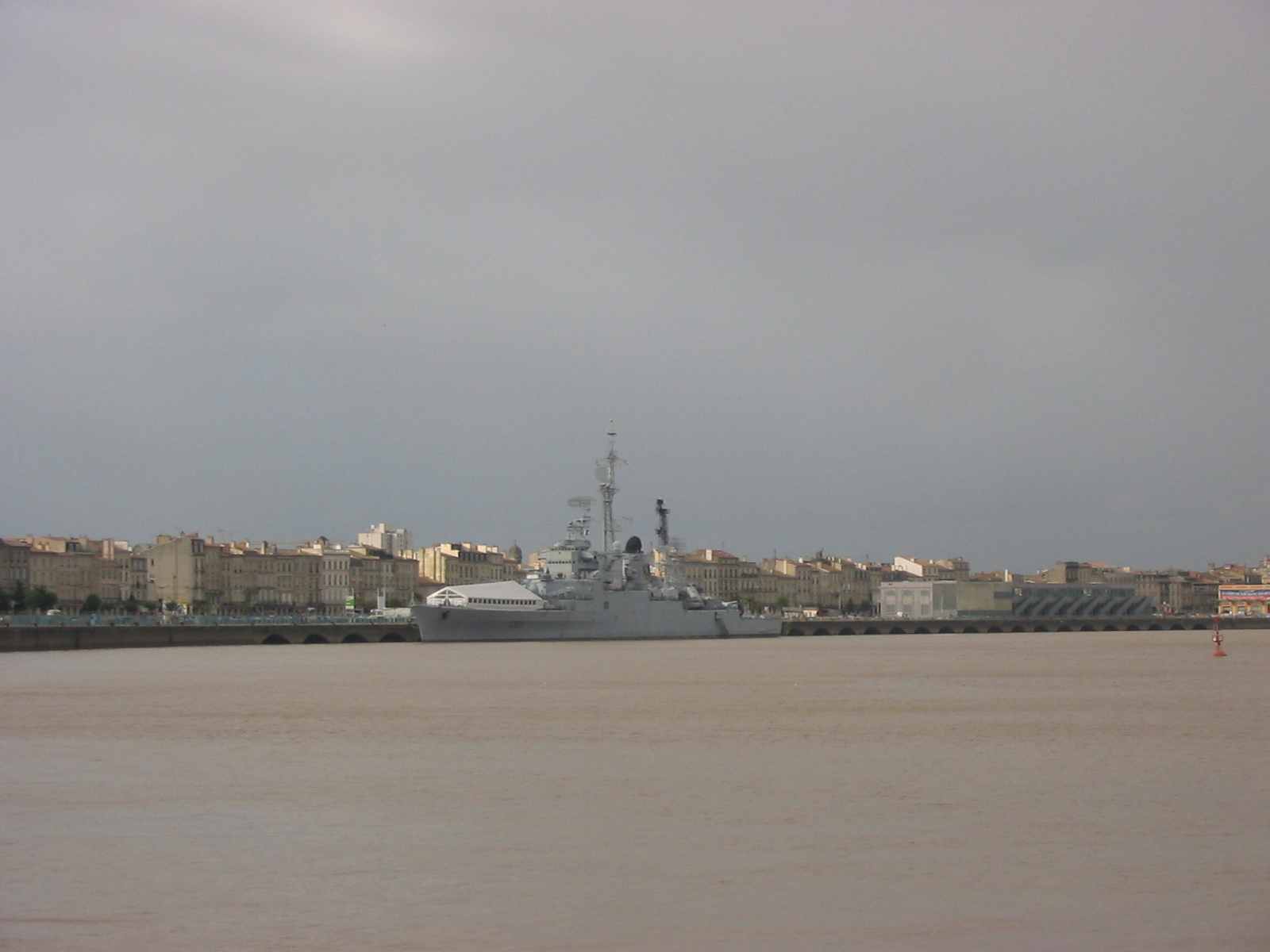}}%
\hspace{8pt}%
\subfigure[][]{%
\label{fig:exampleCopyDays-d}%
\includegraphics[width=0.16\copydayslength]{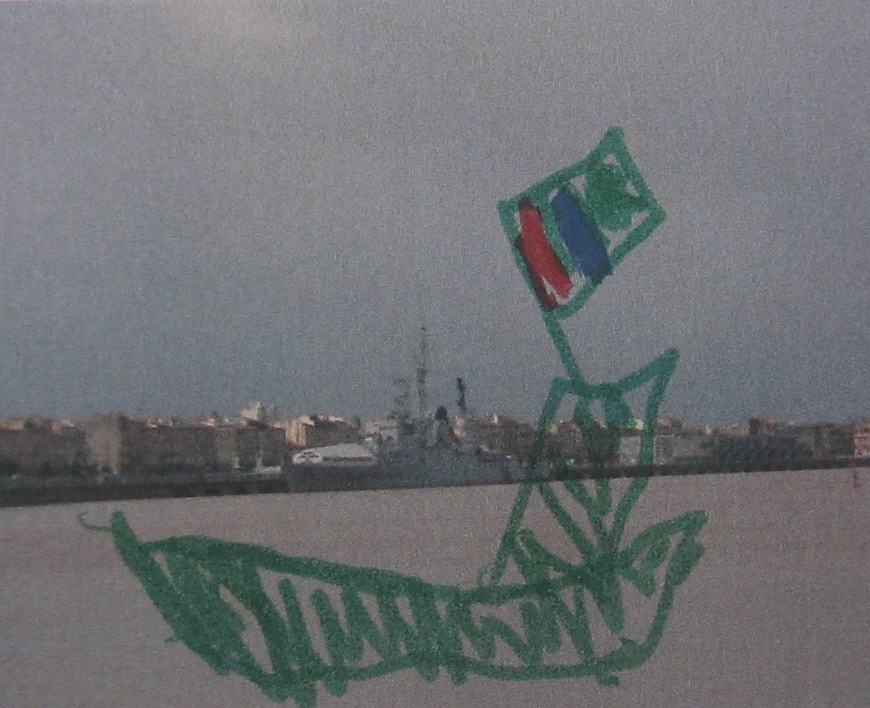}}%
\caption[Examples from Copydays.]{Examples from Copydays:
  \subref{fig:exampleCopyDays-a} and \subref{fig:exampleCopyDays-c}
  are two original images, while \subref{fig:exampleCopyDays-b}
  and~\subref{fig:exampleCopyDays-d} are two strong variants used as
  queries. Size ratio is preserved.}%
\label{fig:exampleCopyDays}%
\end{figure}

%

\subsection{Result Quality}

\begin{figure}
	\centering
	\includegraphics[width=.8\columnwidth]{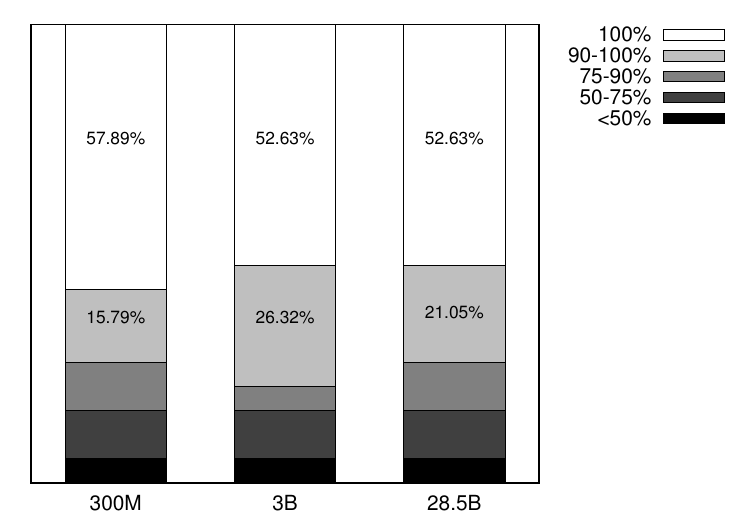}
	\caption{Result quality for the CopyDays benchmark with distracting collections of varying sizes.}
	\label{fig:largeQuality:CD}
\end{figure}

Figure~\ref{fig:largeQuality:CD} gives the result quality for the Copydays image benchmark.  
Overall, the results are excellent for all but the most difficult variants. The \nvtree{} is able to identify the correct images most of the time, even from quite strongly distorted queries. It is not surprising to observe that quality drops with extremely compressed images (a person can sometimes hardly find any similarity between a JPEG 3\% compressed image and its original version) and with some of the strong variants. Note that sometimes such attacked query images create only a handful of vectors, so there are too few matches for the original to rank \#1---it is lost in the noise.

\subsection{Retrieval Performance}

When the three \nvtrees{} fit entirely in main memory, which is the case for the 300M and the 3B collections, then answering each query vector is extremely fast.  A detailed analysis shows that, on average, 2,500 query vectors can be processed per second (the range is from 1,978 to 3,302 vectors per second). On average, therefore, identifying 100 near neighbors of a single query vector takes about 0.4 milliseconds per \nvtree{}. In turn, as there are about 1,000 query descriptors per image, it takes about 400 milliseconds (or 0.4 seconds) to identify the images that are the most similar to the query image. 


When using the 28.5B collection, however, no index fits entirely in main memory. In this case, the system must therefore get data from disks for almost every query vector; as each query vector is likely to access a different part of the index, no main memory buffering policy copes with such a demanding random access pattern. Detailed analysis shows that about 50 query vectors could be processed per second in this case, which is 50 times slower than for the cases where the index fit in RAM. More precisely, in the case of the 28.5B collection, it is possible to return the answer of a query vector in 22.47 milliseconds on average.\footnote{About 2\% of the query vectors could be answered in less than 0.5 millisecond (the data was cached in memory), 91\% of the query vectors could be answered in 20.84 milliseconds on average (with observed times varying from 5ms to 31ms), and 7\% of the query vectors required on average about 50 milliseconds (occasionally read times of over 100 milliseconds were observed). Determining the exact causes of these variations turned out to be extremely complicated because it is difficult to precisely know how the NAS NetApp storage server handles the disk requests. 
} 

Note that since some queries have very few descriptors while others have more descriptors, the retrieval time varies significantly. As pointed out earlier, however, the construction process of the \nvtree{} is such that the search analyses only four leaves per \nvtree{}, and those four leaves are organized such that one disk read is issued. Because three \nvtrees{} are used, no more than three disk reads are performed per descriptor search, and only 3,000 vectors among the 28.5 billion on average, which is about $0.00001\%$ of the collection. The response time of the \nvtree{} is therefore independant of the scale of the collection as soon as memory is outgrown.

\section{Conclusion}
\label{sec:conclusion}

Visual instance search is the task of retrieving from a database of images the ones that contain an instance of a visual query. So far, only local image features are powerful enough to support such fine-grained recognition. Recent progress in approximate high-dimensional indexing has resulted in approaches handling several hundred million to a few billion high-dimensional vectors with excellent response times.  These methods typically rely on residing in main memory for performance.  We argue, however, that data quantity will always win over memory capacity in the long term. Therefore, high-dimensional indexing solutions that are truly concerned with the scalability of the feature collections they manage must address collection sizes beyond RAM capacity and efficiently utilize disks for extending storage.

Furthermore, we argue that scalability is not the only challenge that must be met as high-dimensional indexing methods must also provide dynamicity---the ability to cope with on-line insertions of features into the indexed collection, and durability---the ability to recover from crashes and avoid losing the indexed data if a failure occurs.  As far as we know, no nearest neighbor algorithm published so far is able to cope with all three requirements: scale, dynamicity and durability.

In this article, we have extended an existing disk-based high-dimensional index, the \nvtree{},  such that it enforces 
the ACID properties of transactions. Experiments show that with our implementation dynamic inserts can be efficiently managed: when the index fits in memory, performance is excellent, but when the index no longer fits in memory, performance degrades very gracefully. 

Indeed, the technology described in this article is in use with
one of the main players in the forensics arena with technology deployed at such clients as Interpol. Their search engine is currently able to index and identify videos (at about 40x real time) from a collection of nearly 150 thousand hours of video, and about 700 hours of video material are dynamically inserted to the index every day.

Future directions of research include indexing deep features instead of the hand-crafted SIFT ones, possibly resulting in even higher quality results. Coping with the very high dimensionality of such vectors might suggest to replace some of the random projections with better locality preserving mechanisms such as the ones described in~\cite{DBLP:journals/prl/PauleveJA10}.
\balance
\bibliographystyle{plain}

\end{document}